# Critical Exponents and Particle Multiplicity Distributions in High Energy Collisions

A.Z.Mekjian[1], S.J.Lee[1,2], and T.Csorgo[3]
[1] Department of Physics and Astronomy, Rutgers University, Piscataway, N.J. 08854
[2] Department of Physics, Kyung Hee University, Yongin, KyungGiDo, Korea
[3] MTA KFKI RMKI, H-1525 Budapest 114 PO Box 49, Hungary

Abstract

Data from the L3, Tasso, Opal and Delphi collaborations are analyzed in terms of a statistical model of high energy collisions. The model contains a power law critical exponent $\tau$ and Levy index $\alpha$. These data are used to study values of $\tau$ and $\alpha$. The very high multiplicity events in L3, Opal and Delphi are consistent with a model based on a Feynman-Wilson gas which has a tail exponent $\tau = 3/2$ and $\alpha = 1/2$.

Phenomenological models of hadron multiplicities have been applied to understanding high energy collisions[1-8]. Several of these models are based on specific statistical models of photon count distributions[9,10]. These models were first used as probability distributions for event-by-event fluctuations in high energy collisions in small systems such as $(e^+, e^-)$ and $(e, p)$ in the past. Their use in large systems (heavy ion collisions) is gaining significant attention because of RHIC experiments. For example, large fluctuations in the neutral pion component are a property of a disoriented chiral condensate [11,12] and such fluctuations are being looked for in RHIC experiments. Past interest centered around dynamical theories which would produce large non-poissonian fluctuations seen in experimental data which could be interpreted in terms of fractal behavior and intermittency associated with a possible underlying cascade process. Two specific and frequently used models are the negative binomial model and a distribution based on photon emission from Lorentzian line shapes as initially developed by Glauber [13]. The latter model is also connected to a Feynman-Wilson gas as discussed in detail in ref[14,6] and briefly mentioned below. These two specific models have been extended in a previous paper[4,5,6] and shown to be special cases of a more general model. A main element of this more general model is a power law critical exponent $\tau$ which appears in statistical physics, or a related Levy index $\alpha$ of probability theory. Various values of $\tau$ or $\alpha$ distinguishes different models in our generalized approach which leads to a unified description of count distributions. Some connections of a Levy distribution with Bose-Einstein correlations and HBT features were also discussed in ref[15-17]. A physical interpretation was given in this reference where the fractal properties of a QCD cascade can be measured by a Levy index. In particular, if this cascade of particles arises from a second order QCD phase transition, the Levy index gives the correlation length associated with HBT correlations.

Here we discuss the role of critical exponents and a related Levy index on multiplicity distributions. The determination of critical exponents is a very important aspect of experimental and theoretical physics. For example, it is a dominant theme in condensed matter physics and in theories of phase transitions. Thus an investigation of their role in the shape and properties of particle multiplicity distributions is useful. Data from the L3[18], Tasso[19], Opal[20] and Delphi[21] collaborations are analyzed in terms of this generalized statistical model of high energy collisions to study the best values of the critical exponent $\tau$ of statistical physics or Levy index $\alpha$ of probability theory. These data are precise data with very small error bars over the limited range of multiplicity detected in each experiment. Connections with the phenomena of Bose-Einstein condensation and critical opalescence are also briefly discussed. The Bose-Einstein condensation of ultra cold atoms trapped in laser wells is of great current interest. This phenomenon is useful not only as an example of phase transitions with associated exponents, but it also has played a significant role in a studying of the function of correlations such as in Hanbury-Brown Twiss determinations of the size of the emitting region in heavy ion collisions. Bose-Einstein condensation of pions have also been recently studied in ref.[22].These authors studied particle number fluctuations and showed that the scaled variances of neutral and charged pions increases dramatically near the condensation point. Critical opalescence occurs at the end point of a first order phase transition where a strong enhancement in density fluctuations is seen [23,24]. At this endpoint, the phase transition becomes second order. Similarly, in a heavy ion collision a strong enhancement in particle fluctuations may occur associated with a possible critical point[25]. An exponent $\gamma$, related to $\tau$ can also be used to describe this enhancement in particle fluctuations near a critical point. This exponent was used in statistical physics to describe the singular behavior of the compressibility near a critical point, a phenomenon connected to very large density fluctuations.

We now briefly summarize some theoretical results previous developed regarding our generalized model that will be used. Details can be found in ref.[4,5] so we included only some fundamental defining equations and key factors to make this present paper self contained. Our focus here is to use a generalized model to try to extract an exponent or index from existing data. The count probability distribution for N-particles [1-8] can be obtained from a generating function $G(u)$

$$G(u) = \sum_{N=1}^{\infty} P_N u^N \qquad (1)$$

The $G(u)$ is related to combinants $C_k$ through a connection that reads

$$G(u) = Exp(\sum_{k=1}^{\infty} C_k(u^k - 1)) = Exp(\sum_{k=1}^{\infty} C_k u^k) / Exp(\sum_{k=1}^{\infty} C_k) \qquad (2)$$

The combinants are important for an understanding of the particle correlations. The

$x_k \equiv Exp[C_k]$ is a statistical weight given to a correlated cluster of size k. Van Hove introduced clan variables or Poisson cluster variables to discuss such correlations. The $\Sigma C_k = N_C$ is the number of Poisson clusters or clans, while $n_C = <N>/N_C = \Sigma k C_k / \Sigma C_k$ is the mean number of particles per clan. Moments of the $C_k$ distribution related to the shape of the multiplicity distribution. The mean multiplicity is $<N> = \Sigma k C_k$ which is the first moment of this distribution. The $m$'th order factorial cumulant is defined as $F_m \equiv \Sigma k(k-1)...(k-m+1) C_k$ which can be normalized as: $\kappa_m \equiv F_m / <N>^m$. For $m=2,3$, the $m$'th central moment $<(N-<N>)^m> = \Sigma k^m C_k$. The fluctuation $\delta N^2 = <N^2> - <N>^2 = F_2 + F_1$ is the second moment of the $C_k$ distribution and the skewness $<(N-<N>)^3> = F_3 + 2F_2 + F_1$ is the third moment. The fourth central moment $<(N-<N>)^4>$ satisfies the equation:

$<(N-<N>)^4> - 3<(N-<N>)^2>^2 = \Sigma k^4 C_k = F_4 + 6F_3 + 7F_2 + F_1$. The two frequently discussed cases are the Poisson and negative binomial (NB) distributions. The Poisson distribution is taken as a baseline for comparison. The NB distribution has been used as a standard parametrization for the count probability $P(n)$ when non-poissonian fluctuations are present. The combinants for a negative binomial are given by a $k$-dependent distribution which is $C_k = x z^k / k$. The $<N> = xz/(1-z)$ and $<N^2> - <N>^2 = <N> + B<N>^2$ with $B = 1/x$. If $z \to 1$, $<N> \to \infty$. A value of $z<1$ exponentially truncates the distribution of $C_k$ and leads to finite moments of it. By comparison, the combinants $C_k$ for a Poisson are $C_1 = <N>$, $C_{k \neq 1} = 0$ and, consequently, $F_2 = 0$ and all higher $F_m = 0$ since only the $k=1$ combinant is none zero. The Poisson distribution thus serves as an extreme limit of a single combinant. Generating functions have recently been used for particle multiplicity distributions in field theories coupled to strong time dependent external sources [26], an example being the color glass condensate. Even in the limit where classical approximations are valid, the multiplicity distribution is non-Poissonian. Dremin [27] has developed a simple analytic model of particle production with strong external sources which has features that are associated with a phase transition.

We next briefly discuss a generalized hypergeometric model that contains, as special cases, several models of particle multiplicity distributions and also features associated with Bose-Einstein condensation phenomena and critical opalescence [5] . It will allow us to study various situations in a unified way. The model involves the hypergeometric function $_2F_1(a,b;c,z)$ which appears in evaluating the generating function $G(u)$ for a specific form of $C_k$. The subscripts 2,1 in $_2F_1$ will be omitted to shorten the notation. The most general hypergeometric model will have a combinant structure $C_k$ given by:

$$C_k = x \frac{[a]_{k-1}[b]_{k-1}}{[c]_{k-1}(k-1)!} z^k \qquad (3)$$

The $[a]_k = a(a+1)(a+2)....(a+k-1)$, with the $[a]_0 = 1$. The $C_1 = xz$. The sums that appear in the exponential parts of eq.(2) are given in terms of $F(a,b;c,z)$ as

$$\sum_{k=1}^{\infty} C_k u^k = xzuF(a,b;c,zu) \qquad (4)$$

The $x$ and $z$ are treated as variables in this approach. The $x$ and $z$ determine the mean multiplicity and its fluctuation. This feature was already illustrated for the negative binomial distribution which has $a=1, b=1, c=2$ in the expression for $C_k$ in eq.(3). In this paper we will focus on the restricted case where $(a,b,c)$ are as follows: variable $a$ in the range $0 \le a \le 1$, fixed $b,c$ at $b=1, c=2$ [4]. This particular 3 parameter model encompasses the Poisson distribution when $a \to 0$, the NB distribution when $a \to 1$, and a third distribution when $a \to 1/2$. Specifically, for $a = 1/2$ the associated multiplicity distribution of $F(a,1;2,z)$ is another model taken from quantum optics initially due to Glauber[13] and subsequently applied to particle multiplicity distributions as in ref[6]. The case $a = 1/2, b = 1, c = 2$ has connections with a Feynman/Wilson gas which has critical exponents[6,14]. This connection relates the variable $z$ that appears in $C_k$ to a critical temperature $T_c$ as $z = 1/(1 + \gamma(T - T_c))$ [6] so that $z \to 1$ as $T \to T_c$. One consequence is a divergence of the specific heat which behaves as $1/(T - T_c)^{3/2}$. The critical exponent associated with the divergence of the specific heat is called $\alpha$ and thus $\alpha = 3/2$. Because of these three special connections, Poisson, NB and Feynman-Wilson gas, a study of the case with variable $a$ seems appropriate.  However, we also mention some interesting connections for $a = 1/2$ and also $a = 2/3$ and with fixed $b,c$ at $b = 1, c = 3$ [5]. Specifically, a parallel with Bose-Einstein condensation and its application to multiplicity count distributions[5] has a specific choice for $a,b,c$ which is $a = 1/2, b = 1, c = 3$. A second set with $a = 2/3, b = 1, c = 3$ leads to results that have some connection with the phenomena of critical opalescence[5]. We include such cases for some comparison.

The general model $F(a,b:c,z)$ allows us to study properties of the multiplicity distribution which are connected to different exponents associated with the behavior of the combinants $C_k$. Specifically, the general $F(a,b;c,z)$ has an associated $C_k \to [\Gamma(c)/(\Gamma(a)\Gamma(b))] \cdot xz^k / k^{1+c-a-b}$ as $k \to \infty$ and, consequently an exponent $\tau = 1 + c - a - b$. The exponent $\tau$ in the $1/k^\tau$ behavior of $C_k$ for large $k$ determines the power law fall off of the combinants. This exponent also appears in standard descriptions of Bose-Einstein condensation of non-relativistic ideal Bose gases where it is related to the dimension of the system $d$ as $\tau = 1 + d/2$. Sums over $1/k^\tau$ give the well-known zeta functions in the theory which appear properties of the critical condensation point. For $a=1/2$, the $C_k = (1/k)(2(k-1))!/((k-1)!(k-1)!2^{2(k-1)})xz^k$ which, for large $k$, falls as $C_k \sim xz^k/k^{3/2}$. Then $\tau = 1$ for $a=1$ and $\tau = 3/2$ for $a=1/2$ for the case $F(a,1;2,z)$. Thus, varying the indices $a,b,c$ allows for a study of the $\tau$ dependence of various quantities. For general $a$ and with $b,c$ still restricted by $b=1, c=2$, the $C_k \sim xz^k/k^{2-a}$ for large $k$ so that $\tau = 2-a$. An exponent $\gamma$ can be introduced which describes the divergence of the fluctuation $\delta N^2 \sim 1/(1-z)^\gamma$ as $z \to 1$. Within the most general set $\{a,b,c;x,z\}$ the two subgroups, each with three parameters, are then $\{a,b,c;x,z\} = \{a,1,2;x,z\}$ and $\{a,b,c;x,z\} = \{a,1,3;x,z\}$. We omit the $x,z$ notation. We will call the model based on the subgroup $\{a,1,2;x,z\}$ HGa2 and the model based on $\{a,1,3;x,z\}$ HGa3. Some of the features associated with these two subgroups are as follows. For $\{a,b,c\} = \{a,1,2\}$, the exponents $\tau = 2-a, \gamma = 1+a$ and the associated Levy index $\alpha$ is given by $\alpha = 1-a$. For $\{a,b,c\} = \{a,1,3\}$ the $\tau = 3-a, \gamma = a$. In both cases $\{a,1,2\}$ and $\{a,1,3\}$, we take $a \leq 1$ to have non-negative $\alpha$. The skewness $Sk = <(N-<N>)^3>$ diverges in both subgroups for $z \to 1$ as follows. For $\{a,b,c\} = \{a,1,2\}$, the $Sk \to xa(1+a)/(1-z)^{2+a}$; for $\{a,b,c\} = \{a,1,3\}$, $Sk \to 2xa/(1-z)^{1+a}$. For large $N$, the $P_N \to 1/N^1$ for the NB distribution and $P_N \to 1/N^{3/2}$ for the Feynman-Wilson gas as $z \to 1$. The exponent associated with this power law decrease of $P_N$ is the same as that of the combinant exponent.

 We now present results for the analysis of the L3, Tasso, Opal and Delphi data [18-21] using the hypergeometric model. Both HGa2 and HGa3 have two other variables, $x$ and $z$, besides $a$, giving three unknowns. The $F_1 = <N>$, $\delta N^2$ or $F_2 = <N^2> - <N>^2 - <N>$, and $<(N-<N>^3)$ or the related $F_3$ can be used to determine $\{a,x,z\}$. For the HGa2 case, the set $\{a,x,z\}$ can be determined as follows [4]. For HGa2 the normalized factorial cumulants satisfy the equation

$$\kappa_m/\kappa_2^{m-1} = a^{1-m}\Gamma(a+m-1)/\Gamma(a) \tag{5}$$

and in particular $\kappa_3/\kappa_2^2 = (a+1)/a$ determines $a$. Next $F_2/<N> = az/(1-z)$ determines $z$ and $<N> = xz/(1-z)^a$ gives $x$. The expressions for HGa3 are also analytic but somewhat complicated and will not be given [5]. A numerical solution for the set $\{a,x,z\}$ is still determined by combinations of $F_1, F_2, F_3$ or $Sk = F_3 + 3F_2 + F_1$. Table 1 gives the values of $\{a,x,z\}$ for HGa2 and HGa3 for the four experimental situations considered. The experimental values of $F_1, F_2,$ and $F_3$ that were used to determine $\{a,x,z\}$ are also given. To distinguish the two models considered we consider the fourth central moment $<(N-<N>)^4>$. Values of $F_4$ are then calculated and compared with data. Truncated values of $F_4$ are also given since the experimental data has a maximum multiplicity $N$.

**Table 1** The set $\{a, x, z\}$ for HGa2 and HGa3 and the associated value of $F_m$

|  | $F_1$ | $F_2$ | $F_3$ | $F_4$ | $a$ | $x$ | $z$ |
|---|---|---|---|---|---|---|---|
| Delphi | 10.6289 | 7.7484 | 26.9274 | 63.5664 | | | |
| HGa2 | 10.6289 | 7.7484 | 26.9274 | 167.5276 | 0.26545 | 10.2114 | 0.73306 |
| truncated | 10.6231 | 7.6417 | 24.4967 | 111.2252 | | | |
| HGa2 | 10.6289 | 7.7484 | 16.9456 | 61.7659 | 1/2 | 11.4294 | 0.59316 |
| truncated | 10.6270 | 7.7154 | 16.2310 | 46.0973 | | | |
| HGa3 | 10.6289 | 7.7481 | 26.9241 | 188.1612 | 0.41500 | 9.09578 | 0.80649 |
| truncated | 10.6219 | 7.6167 | 23.8494 | 114.4277 | | | |
| HGa3 | 10.6289 | 7.7486 | 22.6234 | 127.1120 | 1/2 | 9.35609 | 0.75977 |
| truncated | 10.6245 | 7.6691 | 20.8257 | 85.7017 | | | |
| | | | | | | | |
| Opal | 10.1049 | 8.3316 | 30.3448 | 89.4123 | | | |
| HGa2 | 10.1049 | 8.3316 | 30.3448 | 196.0200 | 0.29263 | 9.25108 | 0.73806 |
| truncated | 10.1018 | 8.2668 | 28.6796 | 153.0886 | | | |
| HGa2 | 10.1049 | 8.3316 | 20.6085 | 84.9596 | 1/2 | 9.97348 | 0.62250 |
| truncated | 10.1039 | 8.3128 | 20.1457 | 73.5659 | | | |
| HGa3 | 10.1049 | 8.3305 | 30.3647 | 219.3170 | 0.46000 | 8.25773 | 0.80830 |
| truncated | 10.1009 | 8.2449 | 28.1095 | 159.2251 | | | |
| HGa3 | 10.1049 | 8.3318 | 28.0886 | 183.8368 | 1/2 | 8.33474 | 0.78813 |
| truncated | 10.1018 | 8.2670 | 26.4046 | 139.7890 | | | |
| | | | | | | | |
| L3 | 19.8667 | 17.5164 | 68.6464 | 177.3715 | | | |
| HGa2 | 19.8667 | 17.5164 | 68.6464 | 477.5231 | 0.29029 | 17.61111 | 0.75231 |
| truncated | 19.8648 | 17.4690 | 67.0036 | 418.3497 | | | |
| HGa2 | 19.8667 | 17.5164 | 46.3324 | 204.2553 | 1/2 | 18.72826 | 0.63813 |
| truncated | 19.8661 | 17.5032 | 45.8873 | 188.6968 | | | |
| HGa3 | 19.8667 | 17.5165 | 68.6445 | 534.9115 | 0.46400 | 15.74033 | 0.82022 |
| truncated | 19.8642 | 17.4511 | 66.3289 | 449.5975 | | | |
| HGa3 | 19.8667 | 17.5169 | 63.9238 | 455.2005 | 1/2 | 15.83883 | 0.80272 |
| truncated | 19.8648 | 17.4676 | 62.1962 | 392.3034 | | | |
| Tasso | 13.5817 | 3.5746 | 11.3816 | 41.9824 | | | |
| HGa2 | 13.5817 | 3.5746 | 11.3816 | 69.4835 | 0.09011 | 16.11965 | 0.74495 |
| truncated | 13.5788 | 3.5290 | 10.4010 | 47.3837 | | | |
| HGa2 | 13.5817 | 3.5746 | 2.8224 | 3.7142 | 1/2 | 31.87739 | 0.34486 |
| truncated | 13.5814 | 3.5711 | 2.7540 | 2.3094 | | | |
| HGa3 | 13.5817 | 3.5752 | 11.3822 | 82.8209 | 0.14140 | 14.24601 | 0.83570 |
| truncated | 13.5778 | 3.5116 | 9.9453 | 48.3953 | | | |
| HGa3 | 13.5817 | 3.5745 | 3.3866 | 5.8855 | 1/2 | 23.18398 | 0.47865 |
| truncated | 13.5814 | 3.5703 | 3.3036 | 4.1743 | | | |

Table1 shows that both HGa2 and HGa3 with fitted parameter '$a$' over estimate $F_4$ by a factor of approximately 2, with HGa2 having a slightly smaller $F_4$. HGa2 with parameter $a = 1/2$ has much smaller values of $F_4$ and is more in line with experiment for $F_4$. However, now the $F_3$'s are not in such good agreement with experiment.

Figure1 show plots of $P_N$. The dots are experimental points, while the various types of lines are the results of the four models considered in Table1. All models give a rather good fit to the data and they differ basically in the high multiplicity tail events which have larger relative error bars [18-21]. Error bars are not shown in the figure. The HGa2 model with $a=1/2$ describes some of the high multiplicity events in Delphi, Opal and L3 as can be seen in Figure1. However, the Tasso data is not fit with HGa2 model with $a=1/2$ which can be described better by HGa2 with fitted $a$. The difference in the high $P_N$ region (see insert in L3 figure) is important for $F_m$ while the difference in high multiplicity region is most noticeable in the figure in a log scale.

**Table 2 Weighted and non-weighted values of chi-squared**
The tail part is for the last 10 data points. The other part is for the remaining points.

| | | a | No weight | Weighted | Tail part | Other part |
|---|---|---|---|---|---|---|
| Delphi | HGa2 | 0.26545 | 0.00010 | 133.69745 | 109.70957 | 23.98788 |
| | | 1/2 | 0.00022 | 74.41869 | 10.47836 | 63.94032 |
| | HGa3 | 0.41500 | 0.00012 | 153.38773 | 124.25874 | 29.12899 |
| | | 1/2 | 0.00015 | 91.48960 | 52.90333 | 38.58626 |
| Opal | HGa2 | 0.29263 | 0.00010 | 34.78589 | 13.77188 | 21.01401 |
| | | 1/2 | 0.00022 | 47.98876 | 1.87741 | 46.11135 |
| | HGa3 | 0.46000 | 0.00012 | 43.77767 | 17.73915 | 26.03853 |
| | | 1/2 | 0.00013 | 39.54677 | 11.00892 | 28.53785 |
| L3 | HGa2 | 0.29029 | 0.00059 | 618.68892 | 95.27207 | 523.41685 |
| | | 1/2 | 0.00076 | 649.96767 | 5.88916 | 644.07851 |
|       | HGa3 | 0.46400 | 0.00063 | 681.50002 | 129.64493 | 551.85509 |
|       |      | 1/2     | 0.00065 | 645.57851 |  84.15993 | 561.41858 |
| Tasso | HGa2 | 0.09011 | 0.00116 | 266.98293 | 117.59639 | 149.38654 |
|       |      | 1/2     | 0.00146 | 307.46195 | 165.15005 | 142.31190 |
|       | HGa3 | 0.14140 | 0.00119 | 268.86557 | 118.64019 | 150.22538 |
|       |      | 1/2     | 0.00143 | 302.60985 | 160.90374 | 141.70611 |

Table 2 gives values for chi squared for the various cases considered. The chi-square values are evaluated using $\Sigma W*(\Delta P_N)^2$, where $W$ is a weight factor, $\Delta P_N = (P_N^{th} - P_N^{exp})$ is the difference between the model prediction and the experimental multiplicity for $N$ particles and the sum is over the experimentally observed particles. Both weighted (with an inverse square error bar weight) and non weighted ($W = 1$) chi-square values are presented. From the values of chi squared we observe the following. The best overall model based on a chi-squared analysis is HGa2 with associated exponents $\tau$ less than 2 rather than HGa3. The case of HGa3 has an associated $\tau > 2$ and this higher $\tau$ over predicts high multiplicity events as can be seen in Fig1. The un-weighted chi-squares for all four data sets are smallest for HGa2 compared to HGa3 when the parameter $a$ is determined by fitting either model to the first three factorial moments of the distribution. Within the HGa2 models we considered two cases. One case is where the parameter $a$ is the value determined by fitting the first three factorial moments, while the other case has $a$ fixed at ½, the Feynman-Wilson gas value, and the first two factorial moments are fitted to the data. Within these choices, the non-weighted chi-squared analysis favors the case where $a$ is fitted to the data. However, for the weighted chi-squared analysis, Delphi data have a smaller weighted chi-square for $a=1/2$. In this data set most of the weighted chi-squared values come from the tail part of the data set which has relatively small error bars. We have also looked at the tail regions of the other data sets and these are also given in the table. Table 2 shows that Opal, L3 data and Delphi data with $a=1/2$ have the smallest chi-squared. For Tasso data, the fitted $a$ case has a smaller chi-squared for the tail.

In this paper we analyzed particle multiplicity distributions using a generalized model and compared results with several different experimental data. Multiplicity data from Delphi, Opal, L3 and Tasso collaborations were used in this comparison. The generalized model stressed here were called HGa2 and HGa3 which each contained three parameters $\{a, x, z\}$ in the hypergeometric variables of eq.3. Special cases for HGa2 are: 1) $a=1/2$, the Feynman-Wilson gas in particle production or Glauber model in quantum optics; 2) $a=1$ the negative binomial model; 3) $a \rightarrow 0$ limit, the Poisson distribution. For HGa3, the choice $a=1/2$ parallels Bose-Einstein condensation phenomena. A critical exponent $\tau$ is linked to the parameter $a$ as $\tau = 2 - a$ for HGa2 and $\tau = 3 - a$ for HGa3. Critical exponents play an important role in various areas of physics and are used to determine the universality class of a model. The factorial moments $F_1, F_2, F_3$ and $F_4$ of the distribution were used in our study. We also did a weighted and non-weighted chi-squared study of the data. The results of our study were given in Tables 1 and 2 and fits to the data are shown in Fig.1. As discussed, the data is better described by HGa2 as compared to HGa3. This result leads us to conclude that $\tau < 2$. By comparison, thermal phase transitions [23] such as a liquid gas phase transition or an Ising model have $\tau > 2$. One physical reason for this difference is that in thermal phase transitions the mean number of particles $<N> = \sum_k C_k$ is conserved. Here $C_k$ is now the weight for clusters of size $k$. This restricts $\tau > 2$ at a critical point where $C_k$ fall as a pure power law. A physical consequence of a pure power law behavior is the lack of a length scale and the appearance of self-similar behavior. In both particle production models and thermal transitions, fluctuations or variances can become infinite at a critical point. Studies of fluctuations have played an important role in the analysis of experimental data. In a liquid gas phase transition fluctuation give rise to the phenomena of critical opalescence. Within the HGa2 model, the special case $a = 1/2$ (Feynman-Wilson gas) has the interesting feature of describing the very high multiplicity events in three of the four experiments considered - Delphi, Opal and L3 as seen in Fig1. The Feynman-Wilson gas has a tail exponent $\tau = 3/2$. Our results also show that the frequently used negative binomial does not give the best description to the data. The negative binomial has a tail exponent equal to 1. We should also mention that studies of the variance have played a dominant and important role in the analysis of experimental data, but more stringent tests may come from higher moments of the distribution. Accounting for high factorial moments represents a challenge to theories. Also of importance is an investigation of the high multiplicity events to extract the tail exponent as shown in fig.1.

This research supported in part by a U.S. DOE Grant No. DE-FG02-96ER-40987. S.J.L. was on sabbatical leave from Kyung Hee University and spent a sabbatical year at Rutgers University in 2006-2007.

Figure caption

Fig.1 : Multiplicity distribution. Solid lines are for HGa2 with fitted $a$ values, dashed lines for HGa3 with fitted $a$ values, dash-dotted lines for HGa2 with $a = 1/2$, and dash-dot-dot-dotted lines for HGa3 with $a = 1/2$ which are very close to the solid line except for Tasso data where this line nearly overlapped with dash-dotted line.

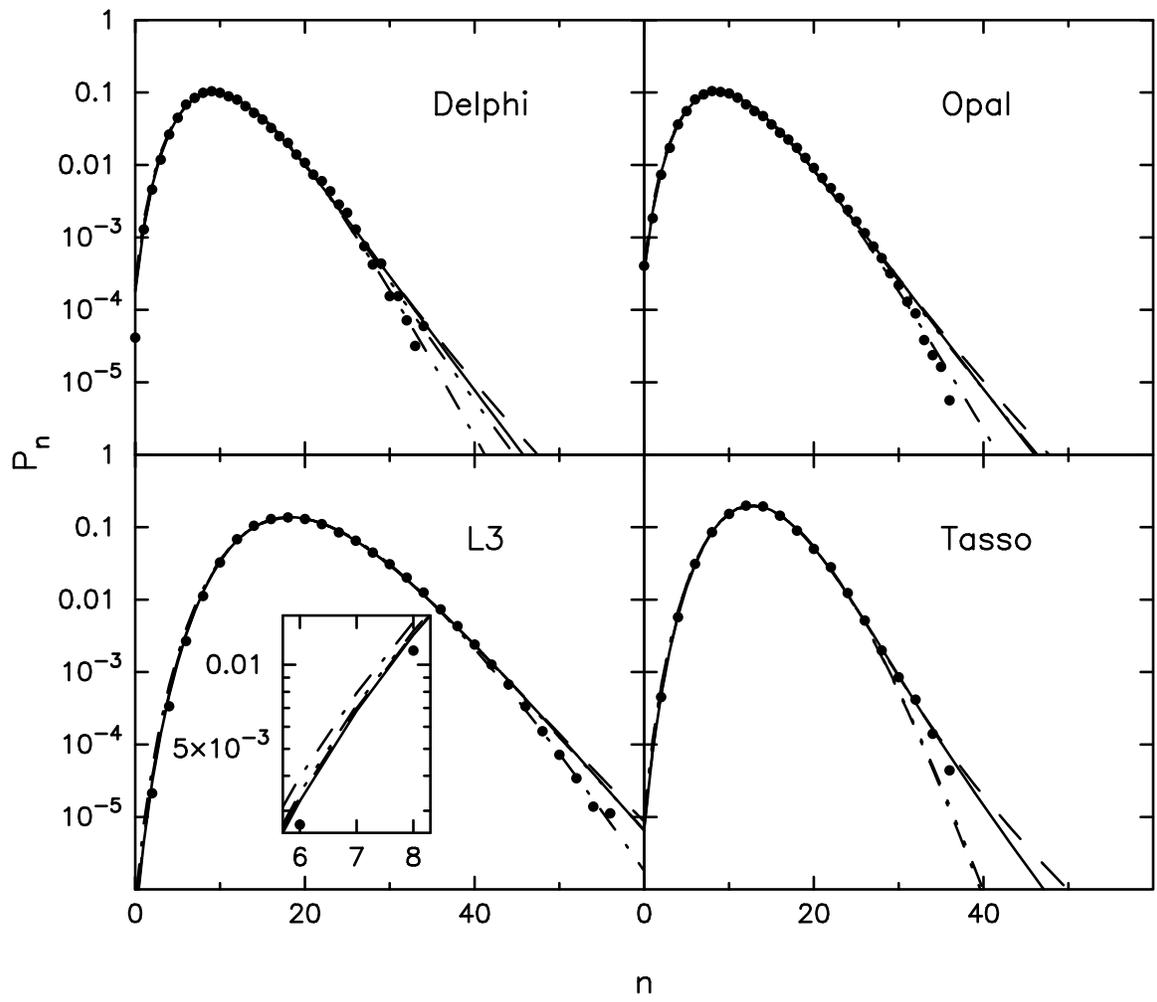